# Causes of evolutionary divergence in prostate cancer

Emre Esentürk[1], Atef Sahli[2], Valeriia Haberland[3], Aleksandra Ziuboniewicz[4], Christopher Wirth[2], G. Steven Bova[5], Robert G Bristow[6,7,8], Mark N Brook[9], Benedikt Brors[10,11,12], Adam Butler[13], Géraldine Cancel-Tassin[14,15], Kevin CL Cheng[16,17], Colin S Cooper[3], Niall M Corcoran[18,19,20], Olivier Cussenot[14], Ros A Eeles[9,21], Francesco Favero[22,23], Clarissa Gerhauser[24], Abraham Gihawi[3], Etsehiwot G Girma[22,23], Vincent J Gnanapragasam[25], Andreas J Gruber[26], Anis Hamid[19], Vanessa M Hayes[27,28,29], Housheng Hansen He[30], Christopher M Hovens[31], Eddie Luidy Imada[32], G. Maria Jakobsdottir[33,34], Chol-hee Jung[35], Francesca Khani[32], Zsofia Kote-Jarai[9,21], Philippe Lamy[36,37], Gregory Leeman[13], Massimo Loda[32,4], Pavlo Lutsik[24,38], Luigi Marchionni[32], Ramyar Molania[39,40], Anthony T Papenfuss[39,40], Diogo Pellegrina[16], Bernard Pope[35,41,19], Lucio R Queiroz[32], Tobias Rausch[42], Jüri Reimand[16,43,17], Brian Robinson[32], Thorsten Schlomm[44], Karina D Sørensen[36,37], Sebastian Uhrig[10], Joachim Weischenfeldt[22,23,44], Yaobo Xu[13], Takafumi N Yamaguchi[45], Claudio Zanettini[32], Andy G Lynch[46], David C Wedge[2], Daniel S Brewer[47,48], Dan J Woodcock[4]

[1]Nuffield Department of Medicine, University of Oxford, UK,
[2]Manchester Cancer Research Centre, The University of Manchester, UK,
[3]Norwich Medical School, University of East Anglia, UK,
[4]Nuffield Department of Surgical Sciences, University of Oxford, UK,
[5]Prostate Cancer Research Center, Faculty of Medicine and Health Technology, Tampere University, Finland,
[6]Manchester Cancer Research Centre and Cancer Research UK Manchester Institute, The University of Manchester, UK,
[7]Christie NHS Foundation Trust, UK,
[8]Div Cancer Sciences, Faculty of Biology, Medicine & Health, University of Manchester, UK,
[9]The Institute of Cancer Research, UK,
[10]Division Applied Bioinformatics, German Cancer Research Center (DKFZ), Germany,
[11]German Cancer Consortium (DKTK), Core Center Heidelberg, and National Center for Tumor Diseases (NCT), Germany,
[12]Medical Faculty Heidelberg and Faculty of Biosciences, Heidelberg University, Germany,
[13]Wellcome Sanger Institute, UK,
[14]CeRePP (Centre de Recherche sur les Pathologies Prostatiques et Urologiques), France,
[15]Sorbonne Université, GRC n°5 Predictive Onco-Urology, APHP, Tenon Hospital, France,
[16]Computational Biology Program, Ontario Institute for Cancer Research, Canada,
[17]Department of Medical Biophysics, University of Toronto, Canada,
[18]Department of Urology, Royal Melbourne Hospital, Australia,
[19]Department of Surgery, The University of Melbourne, Australia,
[20]Department of Urology, Western Health, Australia,
[21]The Royal Marsden NHS Foundation Trust, London & Sutton,
[22]Biotech Research & Innovation Centre (BRIC) - University of Copenhagen, Denmark,
[23]Finsen Laboratory, Copenhagen University Hospital - Rigshospitalet, Denmark,
[24]Division Cancer Epigenomics, German Cancer Research Center (DKFZ), Germany,
[25]Department of Surgery, Urology, University of Cambridge & Cambridge University Hospitals NHS Trust, Cambridge Urology Translational Research and Clinical Trials (Office), Cambridge Biomedical Campus, Addenbrooke's Hospital Site, S Wards Building, UK,

[26]University of Konstanz, Germany,
[27]School of Medical Sciences, University of Sydney, Faculty of Medicine & Health, Australia,
[28]School of Health Systems & Public Health, University of Pretoria, South Africa,
[29]Manchester Cancer Research Centre, University of Manchester, UK,
[30]Princess Margaret Cancer Centre, University Health Network; Department of Medical Biophysics, University of Toronto, Canada,
[31]Department of Surgery, The Collaborative Centre for Genomic Cancer Medicine, The University of Melbourne, Australia,
[32]Department of Pathology and Laboratory Medicine, Weill Cornell Medical College, USA,
[33]Division of Cancer Sciences, The University of Manchester, UK,
[34]Christie Hospital, The Christie NHS Foundation Trust, Manchester Academic Health Science Centre, UK,
[35]Melbourne Bioinformatics, The University of Melbourne, Australia,
[36]Department of Molecular Medicine, Aarhus University Hospital, Denmark,
[37]Department of Clinical Medicine, Aarhus University, Denmark,
[38]Department of Oncology, KU Leuven, Belgium,
[39]Walter and Eliza Hall Institute of Medical Research, Australia,
[40]Department of Medical Biology, The University of Melbourne, Australia,
[41]Australian BioCommons, The University of Melbourne, Australia,
[42]Genome Biology, European Molecular Biology Laboratory (EMBL), Germany,
[43]Department of Molecular Genetics, University of Toronto, Canada,
[44]Department of Urology, Charité - Universitätsmedizin Berlin, Germany,
[45]Jonsson Comprehensive Cancer Center, University of California - Los Angeles, USA,
[46]School of Medicine, School of Mathematics & Statistics, University of St Andrews, UK,
[47]Metabolic Health research centre, Norwich Medical School, University of East Anglia, UK,
[48]The Earlham Institute, UK

*Corresponding authors (email: emre.esenturk@eng.ox.ac.uk, dan.woodcock@nds.ox.ac.uk)

## ABSTRACT

Cancer progression involves the sequential accumulation of genetic alterations that cumulatively shape the tumour phenotype. In prostate cancer, tumours can follow divergent evolutionary trajectories that lead to distinct subtypes, but the causes of this divergence remain unclear. While causal inference could elucidate the factors involved, conventional methods are unsuitable due to the possibility of unobserved confounders and ambiguity in the direction of causality. Here, we propose a method that circumvents these issues and apply it to genomic data from 829 prostate cancer patients. We identify several genetic alterations that drive divergence as well as others that prevent this transition, locking tumours into one trajectory. Further analysis reveals that these genetic alterations may cause each other, implying a positive-feedback loop that accelerates divergence. Our findings provide insights into how cancer subtypes emerge and offer a foundation for genomic surveillance strategies aimed at monitoring the progression of prostate cancer.

Key words: causality, causal inference, cancer evolution, evolutionary divergence, prostate cancer

## INTRODUCTION

Most cancers evolve in a similar way to species and are subject to the same core evolutionary principles of mutation, natural selection and genetic drift[1,2]. One of the key evolutionary mechanisms in the natural world is divergent evolution, which can ultimately lead to the emergence of new distinct species from a common ancestor in a process known as speciation[3]. In the context of cancer, divergent evolution can happen within an individual tumour, leading to distinct subgroups of cancer cells with different properties, such as metastatic potential[4] or resistance to treatment[5]. Divergent evolution also occurs across individuals, where tumours from similar progenitor cells evolve into genetically and phenotypically distinct cancer subtypes, paralleling the concept of speciation in nature.

In prostate cancer (PCa), our recent study[6] revealed two evolutionary pathways known as the canonical and alternative trajectories. To explain how the pathways were related, we proposed the *evotype model of prostate cancer evolution*, in which tumours start on the canonical trajectory but some will diverge onto the alternative trajectory. The cause(s) of this divergence remain unknown, although we found tumours on each pathway. We used these properties to infer the evolutionary trajectory for each tumour and classify them into evolutionary disease types, or *evotypes*, which revealed that Alternative-evotype tumours displayed a worse prognosis than those of the Canonical-evotype[6]. Therefore, understanding the mechanisms that drive this evolutionary divergence would provide valuable insights into the aetiology of these clinically relevant subtypes, aiding clinical decision-making and potentially enabling targeted therapeutic interventions.

The observed association with AR disruption in Alternative-evotype tumours provides clues into the underlying evolutionary dynamics. AR is a central driver of prostate cancer[7-9], and its behaviour changes throughout disease progression, from normal prostate epithelia, to localised tumours, through to metastatic disease[10]. There is evidence that AR behaviour can be disrupted by specific genetic alterations, most notably *CHD1* loss[11-14], which are significantly enriched in Alternative-evotype[6] tumours. Taken together, these findings suggest that certain genetic alterations lead to acquired AR dysregulation, which in turn promotes progression along the alternative trajectory. As direct evidence for this mechanism is lacking, we therefore sought to investigate the causal role of these events in driving divergence in prostate cancer evolution.

Determining causality during cancer evolution is fraught with difficulties. Experimental methods have been developed that can replicate single nucleotide variants[15] but producing copy number alterations (CNAs), which are more commonly observed in prostate cancer[16], is more difficult to perform in the laboratory. Furthermore, there are no established evotype analogues in cell lines, organoids or model organisms making it difficult to relate any experiments back to the human-derived evotypes. As such, the only option is to infer causality from observational data, although this approach presents its own challenges. The standard method for investigating causality involves adjusting for confounders, but in complex systems like cancer evolution, it is not reasonable to assume that all confounders can be measured and controlled for. We therefore need to account for the action of unobserved confounders which requires more nuanced probabilistic approaches such as Mendelian Randomisation[17-18]. However, even these methods generally rely on assumptions that are unlikely to hold in the context of cancer evolution. Recently, alternative treatments of confounders in cancer pathways were also studied using principal components[25] and by inferring the causal network[39] from the data. However, these studies assume linear relationships between variables which may not generally hold.

We therefore propose a semi-parametric two-step method to infer causal relationships in cancer evolution from cross-sectional data. We first provide a non-technical introduction to causal inference, discuss why current methods are not suitable for this application and describe how our method can resolve these issues. We then apply our inference method to CNAs identified in prostate tumours from 829 patients from the Pan Prostate Cancer Group (PPCG) data set[19]. The PPCG has assembled a clinically well annotated and diverse collection of WGS data from 1,209 fresh frozen prostate samples from 1,001 patients. WGS data has been reanalysed with standardised pipelines, annotating high-accuracy somatic variant calls, driver mutations and mutational signatures. Here, we use this dataset to first investigate causal effects in the chain of events from loss of the *CHD1* gene, through AR dysregulation, to evolution along the alternative trajectory. We then refine our approach to incorporate the direction of causality, allowing us to investigate the causal effect of *CHD1* loss on the fixation of other genetic alterations. Next, we show how our method can be used to identify other possible genetic alterations that may also induce AR disruption and therefore cause divergence toward the alternative trajectory. Finally, we discuss the implications of our findings.

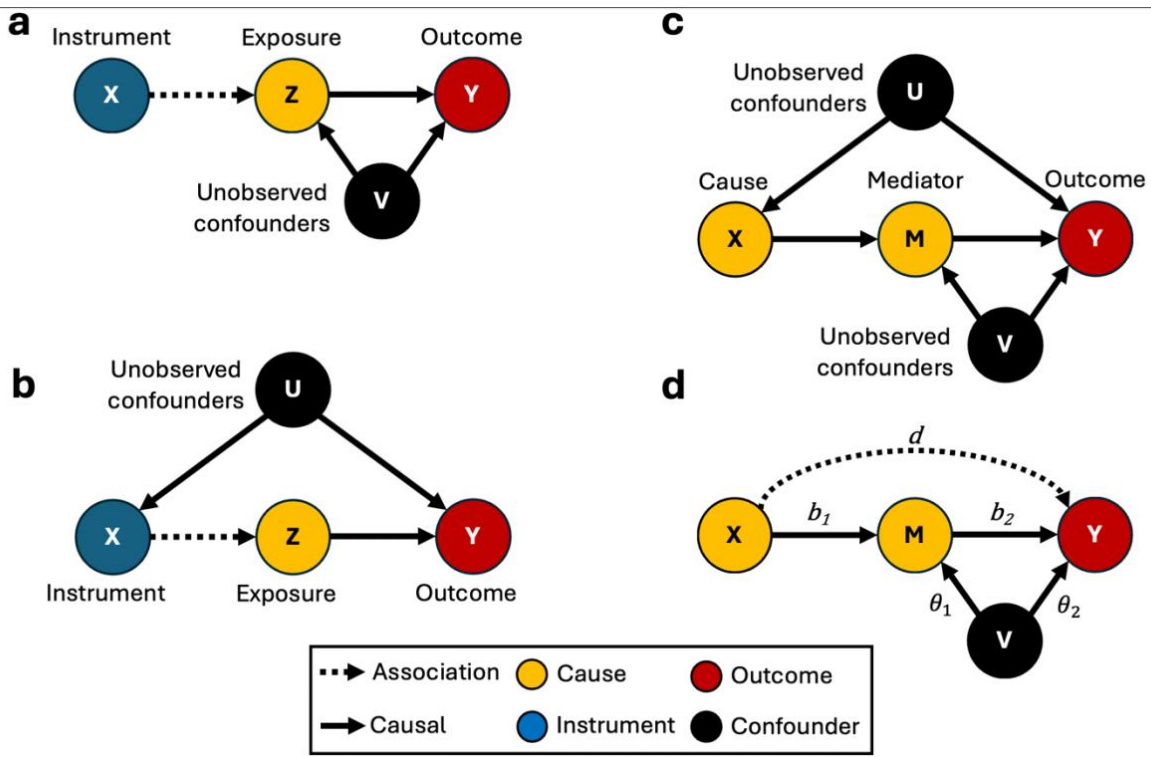

*Figure 1 | Causal inference models* a) The standard instrumental variables/Mendelian Randomisation model that can be used to estimate the causal effect of the exposure on the outcome. b) An invalid instrumental variable model as it has potential confounders that affect the instrument and outcome. c) The proposed causal model that links the cause X and outcome Y, through a mediator M, while incorporating two sets of confounders. d) the effective parametric model after U confounder is subsumed into the network via do-calculus

## RESULTS

### *A two-stage approach for causal inference in cancer evolution*

Approaches for causal inference using observational data are widely applied, but no single method suits every scenario[20-21]. As biological processes are inherently stochastic in nature, we are interested in a probabilistic, rather than deterministic, framing of causality, wherein a cause increases the probability of an outcome rather than guarantying it. In biomedical applications, Mendelian Randomisation (MR) is commonly used to infer causality using inherited single nucleotide polymorphisms (SNPs) as instrumental variables (IVs)[22-24] (Figure 1a). To establish causality, we need to account for unobserved confounders ($V$) that affect both the exposure, $Z$, and the outcome $Y$. MR achieves this by leveraging genetic variants ($X$) that are *associated* with the exposure but not directly causal on the outcome. Since the distribution of SNPs in the population is essentially random with respect to the exposure, a SNP associated with the exposure can be used to replicate the random assignment of participants in a randomised controlled trial, thus allowing us to isolate the effect of the exposure on the outcome.

In principle, MR could therefore be applied to cancer evolution to investigate the causal effects between an exposure, such as AR dysregulation, and an outcome, such as divergence to the alternative trajectory. Indeed, there are a number of somatic genetic alterations that are associated with AR dysregulation as potential instruments instead of SNPs. However, unlike inherited SNPs, these genetic alterations are not randomly distributed in the population as their occurrence and subsequent fixation in the tumour cell population could be influenced by a number of unforeseen external factors (Figure 1b). This violates a key MR assumption, since factors may affect both the presence of these genetic alterations and the likelihood of evolutionary divergence. A more accurate model therefore incorporates both sets of confounders (i.e. merges the graphs in Figures 1a and 1b), but causal inference would be intractable via the MR/IV approach.

We therefore propose a new approach in which the instrument is modified so it is the *cause* ($X$) of some *mediator* ($M$), which itself causes the *outcome* ($Y$). This set up allows us to first account for unobserved confounders $U$ using a causal inference framework known as do-calculus[11], and then estimate the maximal confounding effect of $V$ using a semi-parametric approach. This two-step method therefore allows us to mitigate the effect of both sets of unobserved confounders (Figure 1c) whilst retaining tractability in the calculations. We can use this approach to calculate causal effect sizes between the mediator ($M$) and the outcome ($Y$), mimicking the causal inference performed by MR. Furthermore, we can also quantify the causal effect between cause $X$ and outcome $Y$, mediated by $M$. Performing inference over the entire causal chain provides additional information about the complete process underlying our observations.

We calculate the causal effects through a metric known as the Average Causal Effect ($ACE$), which quantifies the increase or decrease in the probability of observing the effect when $X$ has occurred compared to when it does not. Using do-calculus, we first expressed the $ACE$ in terms of conditional probabilities that remove the influence of confounder $U$, denoted as $ACE_U$. As it is not possible to incorporate confounding through $V$ using do-calculus, we instead sought to identify bounds for $ACE_U$ in the presence of high, but plausible, levels of confounding through $V$. If the causal effect remains consistent after considering these bounds, it therefore indicates that the causal association is robust and holds even in the presence of these potential confounders.

We calculated these by setting up a logistic regression model for the system after application of do-calculus to mitigate for $U$ (Figure 1d). We then followed Imbens' approach[26-27] to estimate the parameters of this model inclusive of a high, but still realistic, level of confounding through $V$. In brief, we first performed parameter estimation without confounding through $V$ (i.e. setting $\theta_1 = \theta_2 = 0$), then used these parameter estimates to set plausible values for $\theta_1$ and $\theta_2$ (Methods). We then performed a second parameter estimation for the remaining parameters to get the final estimates for $b_1$, $b_2$ and $d$. We found that the effect of confounding by $V$ differs depending on the signs of the associated parameters. If $\theta_1$ and $\theta_2$ have the same sign, then $V$ acts to diminish the impact of M → Y, giving the bound $ACE_{U,Vs}$. However, if $\theta_1$ and $\theta_2$ have mixed (opposing) signs, $V$ acts to enhance the causal association between $M$ and $Y$, providing the bound $ACE_{U,Vm}(M,Y)$. These provide lower and upper bounds for $ACE_{U,V}$ in the presence of confounding through $V$, although the bound direction depends on the context of the relationships in the causal chain. In practice, we only utilise the bound with the smallest absolute value. If this bound does not cross zero relative to $ACE_U$, we conclude that the causal effect is robust to confounding by $V$. We report both bounds in this manuscript to aid understanding of the outputs of the method. A full explanation of the approach, including what constitutes plausible values for $\theta_1$ and $\theta_2$ are given in Methods.

This allowed us to calculate six causal metrics:
1) The $ACE_U(M,Y)$ between M → Y, if $V$ can be assumed to be negligible
2) The $ACE_U(X,Y)$ of the causal chain X → M → Y, if $V$ can be assumed to be negligible.
3) The $ACE_{U,Vs}(M,Y)$ of the causal chain M → Y with confounding by $V$ as a common cause of $M$ and $Y$
4) The $ACE_{U,Vm}(M,Y)$ of the causal chain M → Y with confounding by $V$ having opposing effects on $M$ and $Y$
5) The $ACE_{U,Vs}(X,Y)$ of the causal chain X → M → Y with confounding by $V$ as a common cause of $M$ and $Y$
6) The $ACE_{U,Vm}(X,Y)$ of the causal chain X → M → Y with confounding by $V$ having opposing effects on $M$ and $Y$

Confidence intervals for these $ACE$ values can be calculated using a bootstrap approach. If these do not span 0 then we say the causal effect is *significant*.

**AR dysregulation and CHD1 loss are causal to the Alternative-evotype**

A fundamental unresolved question in the evotype model is whether acquired AR dysregulation causes divergence toward the alternative trajectory. To investigate this, we identified a genetic alteration that causes AR dysregulation. Of the genes affected by genetic alterations significantly (Fisher exact, $p < 0.05$) associated with the Alternative-evotype[6], we found that loss of *CHD1* is known to disrupt the AR cistrome[11] as the protein is a known co-factor of AR in DNA binding[8]. Furthermore, loss of heterozygosity (LOH), of a DNA segment spanning the *CHD1* gene occurs in approximately 85% of Alternative-evotype tumours. Therefore, there is strong evidence that *CHD1* LOH is causal to AR dysregulation.

Using previously developed algorithms[6] we first extracted binary variables corresponding to *CHD1* LOH ($CHD1$-), AR dysregulation ($AR$), and Alternative-evotype ($ALT$) for 829 patients in the PPCG data set (Methods). We then calculated the causal effect between AR dysregulation and adherence to the Alternative-evotype for a single confounder $U$. This yielded $ACE_U(AR, ALT) = 0.69$ with a confidence interval CI:[0.52, 0.87] (for %95 confidence level), so this result is significant. This can be interpreted as the probability of developing into an Alternative-evotype tumour if AR dysregulation had occurred is 0.69 higher than if it had not. This is a very strong result considering the range of probability lies in [0,1]. When considering the second confounder $V$, this gave $ACE_{U,Vs}(AR, ALT) = 0.42$, with CI:[0.33, 0.60], again showing statistical significance. As this constitutes a lower bound for the causal effect, this is strong evidence that AR is the primary factor that drives evolution toward the alternative trajectory.

We can also use our method to investigate how *CHD1* LOH itself promotes evolution toward the Alternative evotype. Calculating $ACE_{U,Vs}(CHD1-, ALT)$ gave 0.23 (CI:[0.16, 0.30]), also indicating that the full causal chain was significant. If the effect of $V$ were negligible, we find $ACE_U(CHD1-, ALT) = 0.28$, CI:[0.21, 0.36]. This shows that *CHD1* loss can be an initiating factor in the observed evolutionary divergence.

## *CHD1* LOH is causally linked to other Alternative-evotype aberrations

We extended our analysis to investigate whether *CHD1* loss and AR dysregulation drive the fixation of other CNAs, rather than just evolutionary trajectory (i.e. $CHD1\ LOH \rightarrow AR \rightarrow CNA_j$), where *j* is the index of the CNA. One issue with this is that we do not know whether these other CNAs could themselves causally influence *CHD1* loss as do-calculus does not implicitly account for the possibility of *reverse causation*. We therefore expanded the method to incorporate a fundamental principle of causality that *cause must precede effect*, ensuring that the direction of causality is correct in the calculation (Figure 2a).

To determine event order, we utilised the cancer cell fraction (CCF), which quantifies the proportion of tumour cells carrying each genetic alteration. We created a set of rules based on the CCF values of each CNA for each patient and used this to create new binary inputs and outputs that can be input into our method (Methods). The principle is that if the proportion of cells harbouring the first CNA in the model, $CNA_1$, exceeds that of the second, $CNA_2$, (i.e. $CCF(CNA_1) > CCF(CNA_2)$), it suggests that all cells with $CNA_2$ also exhibit $CNA_1$, which is only possible is $CNA_1$ had occurred before $CNA_2$. In this case we set $X = 1$, $Y = 0$ (and vice versa if the CCF relationship is inverted). However, when the CCFs are equal as the order of these events cannot be inferred. If $CCF(CNA_1) = CCF(CNA_2) = 0$, then we set $X = 0$, $Y = 0$ as we have no evidence the events will occur at all, yet alone what order they are in. If $CCF(CNA_1) = CCF(CNA_2) = 1$, then we know both CNAs have occurred but not in what order they occurred. In this case, we calculate the proportion of the population where $CCF(CNA_1) > CCF(CNA_2)$ and assign this proportion (adjusted for AR status (Methods)) of the $CCF(CNA_1) = CCF(CNA_2) = 1$ patients to $X = 1, Y = 1$, with the remaining proportion set to $X = 0, Y = 1$. We can use these new data to calculate a metric we refer to as the *temporal ACE* ($tACE$). A full description is given in Methods.

We calculated the $tACE$ of $CHD1$ loss toward the set of genetic alterations investigated in our previous study[6]. To aid interpretation, we categorised and ordered the various genetic alterations based on whether they are significantly associated (Fisher exact, *p<0.05*) with the Canonical or Alternative-evotypes and hence referred to these as Canonical or Alternative events. By calculating $tACE_U(CHD1-, CNA_j)$ (Figure 2b), we found that *CHD1* loss displayed a significant positive causal effect on LOH:2q, LOH.13q (*RB1, EDNRB*), GAIN:7, GAIN:8q (*MYC*), and GAIN.3q; all of these are Alternative events. In contrast, we observed a *negative* causal effect for some Canonical events following a *CHD1* LOH, particularly LOH:17p (*TP53*) and LOH:21 (the LOH associated with the *TMPRSS2/ERG* fusion) and HD:10q22 (*PTEN*) – we therefore describe these associations as *anti-causal*. Conceptually, we consider this effect as *blocking* divergence to the Canonical pathway. Calculating the lower bound incorporating potential confounding through $V$, $tACE_{U,Vs}(CHD1-, CNA_j)$, revealed that the causal effect of CHD1 LOH on Alternative events is preserved, except for GAIN:8p (Figure 2c). Similarly, for $tACE_{U,Vm}(CHD1-, CNA_j)$, the upper bound for the effect of $V$, showed the anti-causal relationship between CHD1 LOH and Canonical events LOH:17p (*TP53*), LOH:21 (*TMPRSS2/ERG fusion*), and HD:10q22 (*PTEN*) still hold after confounding.

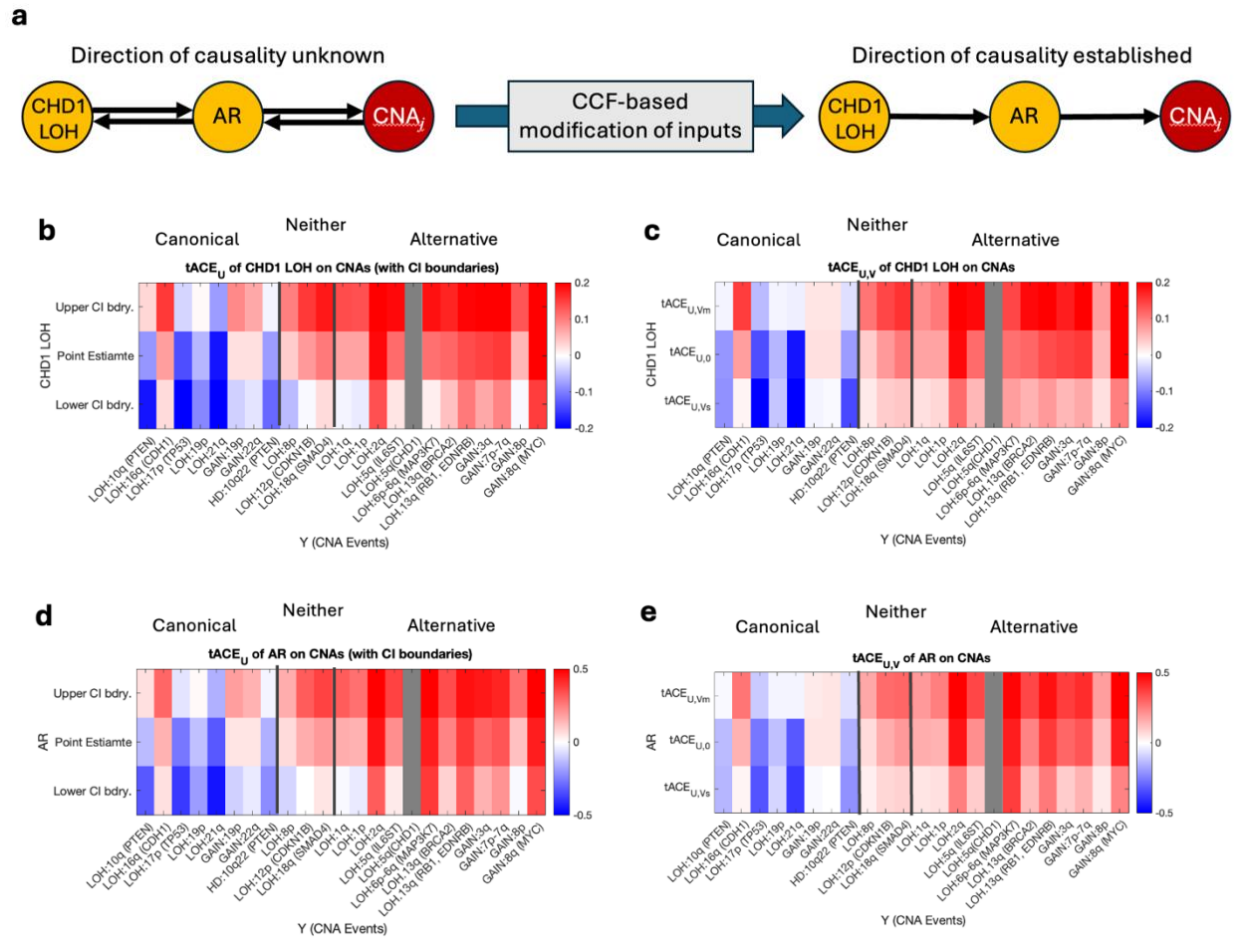

*Figure 2 | (a) A schematic showing how the graph showing the causal chain between CHD1 loss and other genetic alterations is confounded by unknown direction of causality (left); after modifying the inputs with our CCF-based rules we can establish the direction of the causal links between the nodes in this graph (right) (b) heat map of computed $tACE_U(AR, Y)$ (middle row) with upper CI boundary (top row) and lower CI boundary (bottom row)) (c) heat map of computed $tACE_{U,V}(AR, Y)$ with negligible interference of V (middle row); with interference of Vm (top row); with interference of Vs (bottom row) (d) heat map of computed $tACE_U(CHD1-, Y)$ (middle row) with upper CI boundary (top row) and lower CI boundary (bottom row (d) heat map of computed $tACE_{U,V}(CHD1-, Y)$ with negligible interference of V (middle row); with interference of Vm (top row); with interference of Vs (bottom row). The events are displayed in the order of their respective associations with the evotypes - Canonical events to the left followed by events with no significant associations to either evotype in the middle and followed by Alternative events to the right.*

Next, we calculated the direct causal effect of AR dysregulation on these CNAs, $tACE_U(AR, CNA_j)$ (Figure 2d), which revealed a very similar pattern to the results from $tACE_U(CHD1-, CNA_j)$ (Figure 2b). The main difference was that LOH:6q (*MAP3K7*) displayed greater values for the point estimate and the confidence intervals. Similar behaviour was found in the $ACE_{U,V}(AR, CNA_j)$ calculation (Figure 2e). This indicates that while AR dysregulation may act to promote fixation of this genetic alteration, it is not necessarily initiated through *CHD1* loss.

**Alternative-evotype aberrations can cause each other but are blocked by certain CNAs of the Canonical-evotype**

By relaxing this requirement of an established causal link between the cause and the mediator, we can use our method to calculate causal effects for other potential initiating events. This allows us to screen all CNAs and identify those that could themselves be potential causes of subsequent CNAs, mediated through AR dysregulation (i.e., $CNA_i \to AR \to CNA_j$).

We calculated $tACE$ in a pairwise fashion between CNAs, evaluating each CNA's ability to cause others through AR dysregulation (Figure 3). We observe that there are four main quadrants to each subfigure of Figure 3, with the top left and bottom right quadrants showing how CNAs of associated with the same evotype influence each other, and the bottom left and top right quadrants showing how CNAs associated with different evotypes affect each other. The lower right quadrant indicates that Alternative-evotype alterations are generally all causal to each other. The exception to this is when *MAP3K7* LOH or *CHD1* LOH was the second event, in which case they were only caused by each other, as well as 13qLOH (RB1, EDNRB) and 2q LOH. This implies that *CHD1, MAP3K7, EDNRB* and 2q LOH are the primary driving events in divergence toward the Alternative-evotype. Experimental validation of this observation is outside the scope of this study, although there are previous studies that show *MAP3K7* LOH drives enhanced AR signalling which is then amplified through AR cistrome modifications mediated through *CHD1* loss[29-30].

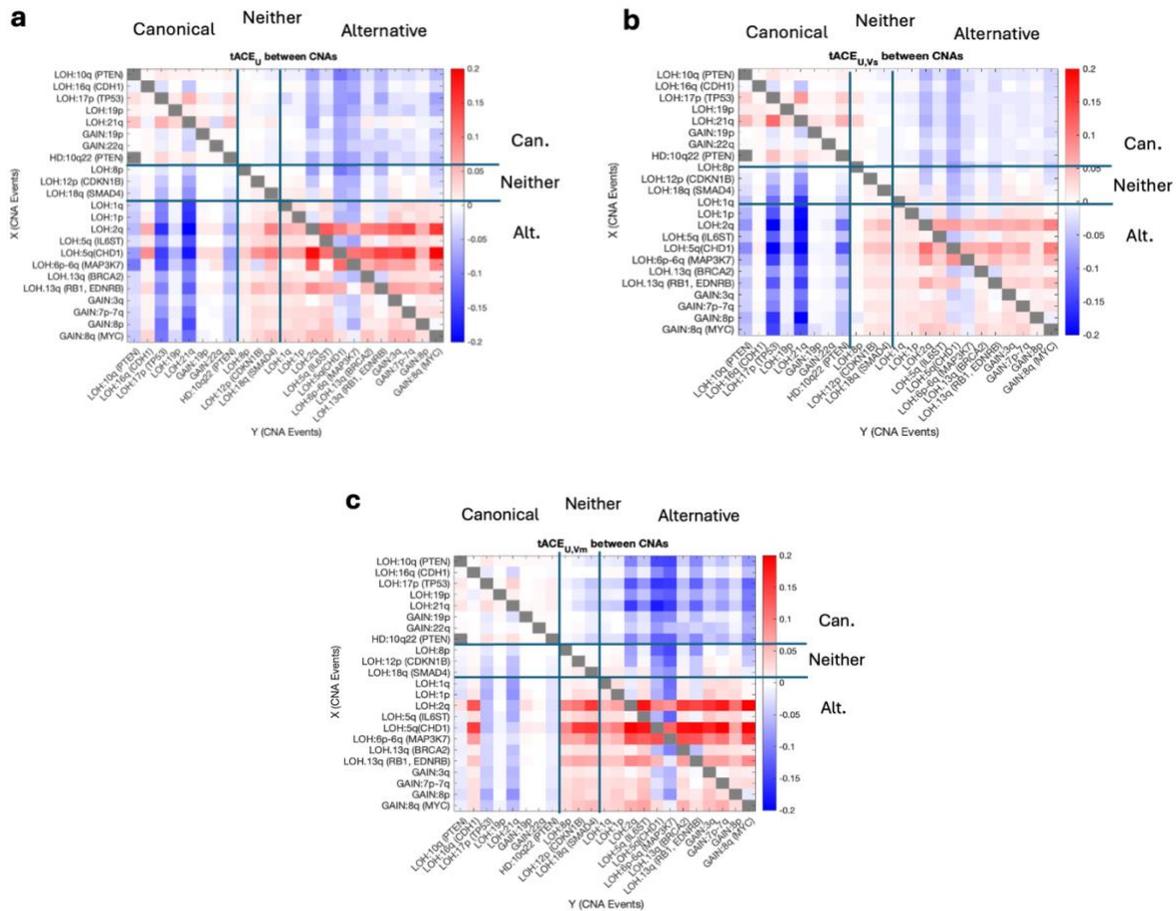

*Figure 3 | Heat map of tACEs between pairs of CNAs. a) temporal causal effect with confounding through U only, $tACE_U(CNA_i, CNA_j)$ (b) temporal causal effect with confounding through Vs set to provide a lower bound for Alternative events, $tACE_{U,Vs}(CNA_i, CNA_j)$. (c) temporal causal effect with confounding through Vm set to provide an upper bound for alternative events, $tACE_{U,Vm}(CNA_i, CNA_j)$. Greyed out associations are not possible (they cannot cause themselves and an LOH always has to precede an HD of the same region).*

The upper left quadrant shows the $tACE$ between genetic alterations of the Canonical-evotype. These generally have a low absolute $tACE$, indicating very weak causal effects through the AR mechanism, if any. Most notably, LOH:17p (*TP53*) and LOH:21q (*TMPRSS2/ERG*) show a positive causal effect to the other Canonical alterations. This counter-intuitive result arises from the fact that the calculations for the ACE and tACE via do-calculus are invariant to whether the binary variable for AR dysregulation is 1 for dysregulated or 1 for tumours with no dysregulation. Therefore, if AR is very rarely dysregulated in these tumours, it implies that it is actually unperturbed AR behaviour that is the link in the causal chain, and this gives the positive tACE value. This could indicate that tumours with TP53 loss and TMPRSS2/ERG fusions actually require normal AR binding, and that AR dysregulation drives tumour cells into an evolutionary dead end so they will die out. This would also explain the mechanism by which they act to "block" the transition to the Alternative-evotype and fixation of other Alternative-evotype associated CNAs.

The lower left quadrant shows the $tACE$ of Canonical-evotype aberrations when an Alternative-evotype aberration occurs first. Interestingly, all of the events are anti-causal to *PTEN*, *TP53* LOH and LOH:21q. This can be understood under our hypothesis that alternative events and AR dysregulation are positively (causally) linked. Then it, follows, from the $tACE_{U,V}(AR, Y)$ calculations in the previous section displayed in Figure 2e that the relationships should be anti-causal. Finally, the top-right quadrant shows that Canonical alterations are anti-causal to almost all Alternative-evotype aberrations, particularly *CHD1, MAP3K7, EDNRB* and 2q LOH – genetic alterations that strongly promote convergence to the Alternative-evotype. This supports the aforementioned behaviour that Canonical-alterations act as "blockers" to the alternative trajectory*.*

**DISCUSSION**

Cancer evolution is a complex process, involving a multitude of interacting factors that may directly affect (cause) the evolutionary trajectory or emerge as a consequence (effect). While associations between genetic alterations and evolutionary outcomes are simple to calculate using significance tests or machine learning feature selection methods, distinguishing causal drivers from downstream effects remains a major challenge.

While the instrumental variable/Mendelian Randomisation method ostensibly provided an appealing framework, it was ultimately not applicable due to potential confounding between the instrument (genetic variant) and the outcome. By reframing the instrumental variable as a causal event, we were able to utilise do-calculus augmented by a semi-parametric modelling approach to estimate causal effects along a three-event causal chain, while accounting for two sources of potential confounders. We treated the effects of confounding between the mediator and the outcome as a realistic "worst-case" scenario, providing three causal effect values: one where this confounding is negligible and lower and upper bounds when it is high. The ability to generate bounds in both directions is critical as the lower and upper bounds allow us to determine if both causal and anti-causal effects respectively are sufficiently strong to persist after confounding. Conceptually, the 'upper bound' is itself an interesting construct as it implicitly models the presence of confounders that are causal to one process but anti-causal to another. This is a departure from many standard causal inference approaches in which the confounder acts as a "common cause" for all the process to which it is linked. The ability to explicitly model confounders with opposing effects is particularly desirable in biological systems, where positive and negative feedback mechanisms often intertwine.

With the inclusion of confidence intervals derived from bootstrapping, our approach therefore provides several layers of information on the robustness of the causal inference. We believe this is the first true causal inference method that can be applied to cancer evolution as methods proposed previously only invoke causality in a loose sense, for instance utilising Suppes' criteria to rule out implausible causal associations rather than estimating the causal effects between nodes directly[31]. The scope of these studies differs to ours in that they aim to identify networks of evolutionary interactions, so it may be possible to integrate our approach into these to provide additional evidence for the network links.

We used our approach to investigate the factors driving evolutionary divergence in prostate cancer evolution using genomic data from 829 patients. Since *CHD1* loss is known to cause AR dysregulation, we designated these as the cause and mediator, respectively. We found that AR behaviour was a statistically significant driver of the evolutionary divergence toward the alternative trajectory, with a probability increase of up to 0.7 in evolving along this pathway when dysregulation had occurred. We believe that this finding constitutes the first robust evidence for a strong causal mechanism underlying the evolution of cancer subtypes in any solid tumour and will inspire future studies using causal inference in cancer evolution. In prostate cancer, it reveals the pivotal role AR plays in driving the evolution of disease subtypes, indicating that AR dysregulation itself could serve as the most informative biomarker in establishing the evolutionary fate and therefore risk profile of a tumour.

We also investigated the evolutionary process at a more granular level, incorporating temporal information into our approach to enable investigation of the causal effects of *CHD1* loss and subsequent AR dysregulation in the emergence and fixation of other genetic alterations. We identified both positive and negative ("anti-") causal effects (Figure 2), indicating that the behaviour of AR can serve to either enhance or inhibit the fixation of other genetic alterations. This indicates the development of a cellular or tumour microenvironment that is (un)favourable to the fixation of other genetic alterations. These findings were borne out when we allowed the causal association between the first two events in the chain to be assumed rather than evidenced (Figure 3). Observing anti-causal associations between genetic alterations associated with different evotypes indicate that evolutionary progression can be blocked to some degree, suggesting that evolutionary steering via therapeutic interventions may be possible to guide the evolutionary progression to a more favourable outcome. This anti-causality captures known underlying exclusive relationships; for instance, the synthetic essentiality of CHD1 in PTEN-deficient tumours[12] manifests as *PTEN* LOH exhibiting the strongest anti-causal effect observed when *CHD1* LOH was treated as the outcome (Figure 3a).

The pairwise analysis between genetic alterations also revealed differences in the evolutionary dynamics of tumours in the canonical and alternative trajectories. In particular, we found that Alternative-evotype-associated alterations were generally positively causal to each other, implying the presence of a positive-feedback loop in which the probability of adherence to the alternative trajectory exponentially increases with the number of these genetic alterations. In contrast, the genetic alterations associated with the Canonical-evotype were causally neutral to each other, implying that locking a tumour onto the canonical trajectory is dependent on the stochastic acquisition of a few key genetic alterations. This information could be used to inform risk-stratification models.

Although our analysis has provided new insights into the evolutionary dynamics of prostate cancer, there are limitations to the method. Any model is a necessary simplification of the underlying processes, and our approach is no exception. By reducing our analysis to a three-step causal chain, we enhance interpretability and tractability, but this simplification may overlook other potential interactions between genetic alterations. To mitigate such model misspecification, we restricted the initial analysis to a well-established causal link between *CHD1* loss and AR dysregulation and incorporated temporal information to ensure the correct direction of causality. We therefore believe that our approach, which mimics individual interventions in an experimental system, provides a level of evidence comparable to other reductionist methods that only focus on relationships that may be part of bigger networks, such as Mendelian Randomisation.

**METHODS**

**Data**

We used whole-genome sequencing (WGS) data from the Pan Prostate Cancer Group (PPCG), comprising an original set of 1,172 samples from 1,001 prostate cancer donors. Further quality control and filtering of this dataset based on sample quality control, copy-number quality and blacklisting is described in a companion manuscript10. Alignment and QC, SNV, indel, CNA and SV calling were performed as described in a companion manuscript. Driver somatic alterations and mutational signatures were also available for this cohort and are reported in two companion manuscripts[34-35].

Using established methodology[36] we determined, for each tumour, which mutations were shared across all cancer cells (hereafter described as 'clonal'), hence defining the MRCA, and which mutations were present in just a fraction of cancer cells ('subclonal'). Clusters of mutations appearing at similar cancer cell fraction[36] (CCF) were used to define subclones. As subclonal mutations necessarily occur after clonal mutations, they may be used to interrogate the temporal dynamics of prostate cancer evolution and mutational processes. To ensure sensitive detection of subclonal mutations in this study, we limited our analysis to prostate cancers that passed all filtering criteria and with a minimum coverage of at least 10 reads per chromosome copy, as previously described[37].

**Copy number alteration calls.** CNAs were called by the Battenberg algorithm[38] and categorised into LOH, HD and Gain. Binary variables denote if the CNA was present of absent in each tumour.
**AR dysregulation.** – DNA double strand breaks (DSBs) can be caused by AR DNA binding[8,9]. We can therefore infer the behaviour of AR by observing the frequency of DSBs proximal to AR binding sites (ARBS). If AR dysregulation had occurred, then the frequency of ARBS-proximal DSBs was significantly lower than if they had occurred at random. We calculated this with a permutation test and output a binary variable to signify if significance had been reached or not.
**Evotype.** The evotypes were established using the approach developed previously[6] and presented as a binary variable of belonging to the Alternative-evotype or not.

**Background on Causality and Inference**

Causal inference is a cornerstone of scientific research, particularly in fields like epidemiology, social sciences, and increasingly in molecular biology and genetics. At its core, causal inference seeks to understand not just the associations or correlations between variables, but the underlying causal relationships that drive these associations. In the context of prostate cancer research, particularly in studying the Alternative-evotype, causal inference allows us to discern the relationships between genetic alterations (CNAs), disruptions in androgen receptor (AR) binding, and subsequent changes in the cellular environment.

**Causation Flow and Directed Acyclic Graphs (DAGs)**

The relationships between various variables can be effectively visualized using Directed Acyclic Graphs (DAGs). A DAG is a type of graph comprising nodes (representing variables) and directed edges (illustrating the directional influence between these variables). These edges in the DAG articulate the conditional (in)dependence structure that exists within the data, mirroring the probabilistic relationships among the variables.

In DAGs, each node symbolizes a distinct variable, connected by directed edges, denoted by arrows (→). The directional nature of these connections is crucial; for instance, if one has $A \rightarrow B$, then 'A' is considered a parent of 'B', and 'B' a child of 'A'. A vital characteristic of DAGs is the absence of cycles – no node is an ancestor or descendant of itself, eliminating any sequence of directed edges that circles back to the originating node. This acyclic nature ensures clarity in the representation of conditional independencies within the graph. A second assumption throughout the paper is the so-called Markov assumption which, in essence means that a node in a DAG is independent of all its non-descendants when conditioned on its parents. Then a DAG not only visually represents these relationships but also admits an associated joint probability density function that aligns with the conditional independence restrictions imposed by the graph's structure. For instance, in a toy example given by A→ B, C→A, C→B with Markov Assumption, one would have

$$P(A, B, C) = P(A|C)P(B|A, C)P(C).$$

In the context of prostate cancer, a DAG might represent how a specific CNAs (X) influence AR binding distribution and how these interactions lead to another CNA event (Y).

**Intervention and Identification: The 'Do' Operator and Observational Probabilities**

In the causal framework, the influence of one variable on another is conceptualised through interventionist principles. A variable X is said to have a causal influence on Y if altering X leads to observable changes in Y's distribution.

To clarify the distinction between observational correlation and causal intervention, Pearl introduced the concept of the 'do-operator'. This operator differentiates between the observational distribution $P(Y|X = x)$, representing the process of seeing, and the interventional distribution $P(Y|do(X = x))$ representing the process of doing. While the former describes the likely values of Y when X is observed to be x, the latter predicts the values of Y when X is deliberately set to x, simulating an intervention. For example, '$do(X = x)$' in our prostate cancer application could simulate the effect of setting a specific CNA event X to a particular state.

The remaining task is evaluating the post-interventional distribution in terms of the knowledge available from observational distributions through mathematical manipulations which is known as the identification, a fundamental aspect of the causal inference. When we can transform an interventional expression (with the 'do' operator) into an observational one (without the 'do'), the expression is said to be identifiable.

**Confounding Effects**

One of the challenges in causal inference is dealing with confounding variables. A confounder (U) is a variable that influences both the cause and the effect, potentially leading to biased or spurious associations if not appropriately accounted for. In prostate cancer research, confounders could include genetic predispositions, environmental factors, or other molecular changes not directly part of the primary causal pathway but influencing both CNAs and AR disruption.

Identifying and adjusting for confounders is critical for accurate causal inference. In our study, we propose a basic model (detailed in the Introduction and beginning of Results) which acknowledges the presence of an unobserved confounders (U and V) that influences both initial CNA events and the AR disruption (denoted by $U$) as well as AR distribution and the outcome variable (denoted by $V$). Hence one needs to deal with these confounders using ideas of do-calculus. To distinguish the spurious correlation effects from the real causal effects one works with a probabilistic measure of causal strength, known as the average causal effect, defined, for binary variables as:

$$ACE_{X,Y} = E[Y|do(X = 1)] - E[Y|do(X = 0)] \qquad (1a)$$

As an alternative measure, it will be convenient and useful to present the causal influence via the concept of odds ratio (OR). OR measures the relative change in the odds of an event with respect to the (multiplicative) probability change in X and ranges from zero to infinity. Values larger than unity indicates positive causality. It is particularly useful in breaking down the causal flow between different nodes. The causal odds ratio (COR) for a binary outcome can be defined by

$$COR_{X,Y} = \frac{P[Y=1|do(X=1)]}{P[Y=0|do(X=1)]} \frac{P[Y=0|do(X=0)]}{P[Y=1|do(X=0)]} \qquad (1b)$$

In the next section, we present tractable mathematical expressions for the specific model we propose and evaluate these quantities for various outcomes Y.

**Causal Modelling of AR Pathways in Prostate Cancer**

Here we first describe our model assumptions and provide a base model for analysis $CHD1\ LOH$ and its $AR$ mediated impact on other CNA events where causal effects can be quantitatively estimated. We then consider generalisation of the base model taking into account the timing effect and possible secondary confounders. We note at the outset that we do not claim to have developed the exact model with complete set of causal connections. Indeed, knowing the full details of every possible genetic interaction is neither (currently) accessible nor practical as such a full-scale model would be intractable for mathematical analysis. Rather, we propose a simple model which is capable of capturing the basic features of the AR mechanism.

**The Base (One-Confounder) Causal Model: CHD1 LOH triggered AR Mechanism for CNAs**

Firstly, based on recent experimental evidence, we know that CHD1 loss causes AR disruption[11] Furthermore, previous experimental studies and recent machine learning classifications[6] suggest that there is a strong correlation with CHD1 LOH, AR disruption and an the Alternative-evotype which hints that there may be a causal link from AR to those CNA events which have high prevalence in the Alternative-evotype. This intuition can be stated as the first assumption of the base model

(i) There is a causal link from CHD1 LOH to other CNA events. If it turns out, upon the results of the analysis, that such causal effect is close to zero, this will indicate that either there is no causal link, or the model needs improvement. Secondly, it is intuitively the case that there are no obvious reasons why a CNA event should directly

(ii) affect another CNA event. But we do know that deviation of AR from a random redistribution is associated with a set of CNA events. Therefore, it is plausible that AR has an impact on a range of CNA events. We therefore make the assumption.
(ii) the main causal influence of CHD1 LOH on other CNA events Y is carried solely via AR route and other causal channels from CHD1 LOH to Y are of lower importance to first degree approximation and are ignored. In other words, *the causal effect of CHD1 LOH on other CNA events is fully mediated by AR.* Such a relationship can be symbolically described as $X \rightarrow AR \rightarrow Y$. Hence, extrapolating the findings of Augello et. al[11] to other CNA events, we look at the causality in X-Y relationship.
(iii) There is an unobserved confounder U that is affecting both X and Y (symbolically described as $U \rightarrow X, U \rightarrow Y$).

It is also plausible that there may be some confounding between AR and Y which is denoted by V. Since this work is a first attempt for the AR dysregulation process, at the simplest level, keeping a balance between modelling and the complexity of reality, we first consider the base model where that AR→Y link is not confounded. However, the influence of possible confounding between AR and Y is also investigated as explained further below.

Having laid out the main tenets of our model, we need to carry out mathematics for identifying the net causal influence of event X on Y which we state here[20-21]

$$P(Y|do(X)) = \sum_{x'} P(Y|A, X') P(X'), \quad (2a)$$

$$P(Y|do(X)) = \sum_{a} P(A|X) \sum_{x'} P(Y|A, X') P(X'), \quad (2b)$$

where $X'$ under the summation denotes the first event ($X' = \{0,1\}$) as a dummy variable. The above expressions quantify the causal relationships between AR and Y; X and Y. For the net causal effect, one computes the difference in the average of these interventional probabilities, i.e., the ACEs. Doing these yields, for $ACE_U$ between AR and Y

$$ACE_U(AR, ALT) = E(Y|do(A = 1)) - E(Y|do(A = 0))$$
$$= \sum_{x'} (P(Y = 1|A = 1, X') - P(Y = 1|A = 0, X')) P(X') \quad (3)$$

**Order of CNA Events and Relevance of Time**

In the first part of the Results, the causal effects were calculated between the CHD1 LOH event and Alternative-evotype implicitly assumed that the causal mechanisms were unidirectional, CHD1 LOH → ALT. Between more granular events (between pairs of CNAs) it may be possible to observe Y causing X.

In this section we extend methods of causal identification making use of the additional time information related to the CNA events. To be more specific, we take the information on cancer cell fractions (CCF) as a proxy for time. That is, roughly speaking, given a sample, a higher CCF value indicates earlier occurrence in time and therefore rules out the possibility that the potential 'effect' could have occurred earlier than that. Then, predicated on this principle we show how the timing information can be used to extend the causal inference.

Firstly, timing information can be used to completely disallow events that when the time-ordering of events are certain. However, one still needs a way to deal with cases where there is uncertainty. This can be done by exploiting the information of definitive cases. Let us consider a simple causal mechanism where, initially, both directions of causal flow are allowed, that is, we have $X \rightarrow AR \rightarrow$

$Y$ as well as $Y \to AR \to X$. We can decompose and restructure the data by a set of rules which will respect time ordering. Let $X_i$, $Y_i$ be CNA events for observation $i$

(i) If $CCF(X_i,) > CCF(Y_i)$, then set $X_i = 1, Y_i = 0$.
(ii) If $CCF(X_i,) < CCF(Y_i)$ then set $X_i = 0, Y_i = 1$.
(iii) If $CCF(X_i,) = CCF(Y_i) = 0$ then set $X_i = 0, Y_i = 0$.
(iv) If $CCF(X_i,) = CCF(Y_i) = 1$ then assign this observation to category (i), i.e. set $X_i = 1, Y_i = 0$ with probability of relative occurrence of $CCF(X_i,) > CCF(Y_i)$ or set $X_i = 0, Y_i = 1$ with probability of relative occurrence of $CCF(X_i,) < CCF(Y_i)$ (both cases adjusted by $AR$ status, i.e., by adjusting, for each patient, the assignment according to the relative occurrences for each of the $AR = 0$ and $AR = 1$ classes) .

With this process the causal direction ambiguity is removed, and methods of do-calculus can readily be applied (Equations (2), (3) in particular).

**Two-confounder model**

The previous section focussed on the refinement of the causal relationships on the base model utilising the time information. Here, we further the causal analysis in a different way, by increasing the complexity. The base model assumes that the information flow from M to Y is not disrupted whereas it may be contaminated with extra confounders affecting both M and Y, i.e., a doubly confounded system. Symbolically this introduces a second (unobserved) confounder V → M, V → $Y$. Diagrammatically, the generalised model is depicted in Figure 1c. We then need to assess how inclusion of extra confounding affects the causal relationships between X (the 1st event), M (the mediator) and Y (the 2nd event) in a way such an analysis can be viewed as sensitivity of the system to possible confounding effects on the original model.

Here, we take an approach due to Imbens[26] and Rosenbaum & Rubin[27] for the sensitivity analysis using a semi-parametric model. The main difference is that our system has double confoundedness where interference of the confounder U between X and Y is treated fully non-parametrically and second confounder V is treated parametrically. The goal is to estimate the parameters of the observable variables using maximum likelihood estimator (MLE). The idea is to obtain a form for the maximum likelihood for the observable system by summing out the first confounder $U$, and thereby reduce the full system to a simpler single confounder system. We then relate model variables in the causal graph regressing $Y$ by $M, X, V$ and regressing $M$ by $X, V$. Since all variables are binary, a logistic form will be appropriate

$$P(Y|M, X, V) = sigmoid(dX + b_2 M + \theta_2 V + c_2),$$
$$P(M|X, V) = sigmoid(b_1 X + \theta_1 V + c_1)$$
$$P(Y, M, X, V) = P(X)P(M|V, X)P(Y|M, X\ V)P(V), \quad (4)$$

where $d, b_2, \theta_2, c_2, b_1, \theta_1, c_1$ are model parameters that quantify the strengths of the relationships as shown in Figure 1d. The use of do-calculus essentially transforms confounding by $U$ into a probabilistic association between $X$ and $Y$, which is parameterised by $d$ in this model. The coupling of the $V$ interferes with the maximum likelihood estimates of the observable parameters depending on the strength of $V$ term. When V is negligible one easily arrives at $P(Y, M, X)$, the probability of a single observable instance (X, M, Y triplet). For the general case, however, we need to sum over U, which, now requires more care and also sum over V which is done explicitly given its coupling strengths $\theta_1, \theta_2$. This means that the model parameters $d, b_2, c_2, b_1, c_1$ will be functions of $\theta_1, \theta_2$. Then, one needs plausible ranges for the parameters' strength of the 'interference' term V. A reasonable assumption is as follows. One considers an extreme hypothetical 'bad' scenario where there is no real causal effect of X on AR and AR on Y and the relationship is purely correlational due to V. Such a scenario could be imagined simulating observations if $\theta_1$ was as large as $b_1$ and $\theta_2$ as large as $b_2$. Hence, for our analysis, a plausibly strong strength for the V interference term can be taken as that its coupling parameters do not exceed the parameter strength of the observable variables, that is,

$$|\theta_1| \leq |b_1|, \quad |\theta_2| \leq max(|b_2|,|d|).$$

Then, we regard the original base model as robust if the interference of V does not lead to changes in the conclusions of the causal outcome which can be calculated via do-calculus:

$$P(Y|do(AR)) = \sum_{x,v} P(Y|AR,X,V)P(X) \tag{5a}$$

$$P(Y|do(X)) = \sum_{a,v} P(A|X,V) \sum_{x'} P(Y|A,X',V)P(X') \tag{5b}$$

Lastly, for the Methods, we discuss a point that is partly tangential to our work but still relevant.

**Selection Bias, Causal Odds Ratio and Collapsibility**

It should be noted that the genetic events that have been studied here are not the existence of a CNA or dysregulation of AR, but rather the event of our observing a CNA or our observing dysregulation. Our powers to observe these events are linked to the selection of these cases (patients' symptoms, cancer stage etc.). Furthermore, criteria for selection can vary systematically between cohorts. The latter problem can be mitigated by carrying out the causal analysis for each cohort and verifying that the causal relationships do not change fundamentally from one cohort to another. The former problem concerning the available/observed cases to be outcome dependent can be partially addressed by focusing on the individual causal links between the nodes. In below these questions are investigated.

To quantify the causal strength, we used both ACE and the causal OR. The results for the ACE were presented in the Results. For COR of CHD1 LOH on the Alternative Evotype, based on the full PPCG data, we found $COR_{CHD1-, \ ALT}$ =4.09 (CI:[2.99, 6.38]). We complement this finding by reporting the results from different cohorts.

The PPCG data consists of cohorts from several countries, namely the UK, Germany, Canada and Australia. We therefore evaluated whether there were any systematic differences between cohorts that fundamentally alter the nature of the causal relation by calculating the ACE and COR for each cohort individually. The results are shown in Table 1. All statistics for all cohorts remained significant, although Germany and Australia generally displayed lower values; this could possibly be attributed to sampling differences between the cohorts (Germany - Early Onset, Australia – High Risk, Canada and UK – Low to Intermediate Risk).

*Table 1. ACE and COR results from different cohorts of PPCG*

|  | PPCG | UK | Germany | Canada | Australia |
|---|---|---|---|---|---|
| $ACE_{CHD1-, \ ALT}$ | 0.28 [0.21, 0.36] | 0.40 [0.26, 0.55] | 0.08 [0.02, 0.18] | 0.34 [0.18, 0.45] | 0.12 [0.07, 0.17] |
| $COR_{CHD1-, \ ALT}$ | 4.09 [2.99, 6.38] | 4.47 [2.92, 8.16] | 5.49 [1.89, 45.8] | 3.98 [2.24, 9.20] | 2.33 [1.40, 4.41] |

To further assess the effect of selection bias, we utilise the (conditional) causal odds ratio as the measure of causality and focus on the (more interesting) link $AR \rightarrow ALT$. In mathematical terms, this amounts to estimate $COR_U^C(AR, ALT.)$ where the superscript $C$ subset of our choice (the subset odds ratios are calculated over). In our case, the subset can refer to a select set (i.e., prostate cancer cases from the cohorts that the data is collected from). Thus, the original causal diagram (Figure 1.b) can be amended appropriately with the inclusion of an additional link from the outcome variable, i.e., $ALT \to S$.

Before we proceed, it is appropriate to introduce a concept that comes useful when dealing with subsets of population is and is also of relevance to our work: collapsibility. Collapsibility refers to the idea that it is possible to make valid inferences on a subset of variables after marginalizing over others. In our context, collapsibility is particularly important because it allows to generalise the associations found in a sampled subgroup (such as cancer progressing) to the whole target population. Hence, we focus on odds ratios as a measure of association as they are suitable for outcome-dependent sampling scenarios.

Definition (Collapsibility of Odds Ratios)[32]. For binary variables X and Y, and disjoint sets of variables B and C, the odds ratio $OR_{X,Y}(B,C)$ or $OR_{Y,X}(B,C)$ given B and C is said to be *collapsible over B* if:

$$OR_{X,Y}(B = b, C = c) = OR_{X,Y}(B = b', C = c) = OR_{X,Y}(C = c) \ for \ all \ b \neq b'.$$

This means the odds ratio remains consistent when we collapse over B. The following theorem is useful for inferring causality in outcome dependent sampling.

Theorem (Whittemore, 1978)[33]. For the conditional odds ratio $OR_{X,Y}(B = b, C = c)$ to be collapsible over B, one of the following conditions is sufficient:

(i) $X \amalg S \mid Y, C$
(ii) $Y \amalg S \mid X, C$

These conditions allow for consistent estimation of the odds ratio when collapsing over the variable B.

To recover the true COR from the observed/selected (computed) COR values (which are based on the available PPCG data) one needs to work with carefully chosen subsets (i.e., conditioning). and make use of collapsibility results for OR[14] (see Methods). To that end we condition on the set $C = (S, \ CHD1 \ LOH)$ and hence
$$COR_U^{CHD1-, \ S}(AR, ALT.) = OR_U^{CHD1-, \ S}(AR, ALT.) = \ OR_U^{CHD1-}(AR, ALT.),$$

where the first equality follows from the fact that all back door paths from $AR$ to $ALT$ are blocked by conditioning on $CHD1 \ LOH$. And the second equality is allowed by observing $AR \amalg S \mid CHD1 \ LOH, ALT$. and hence one can collapse the $OR_U^{CHD1-,S}(AR, ALT.)$ over S, thereby untangling the effect of selection (S) from the causal analysis. This shows that selection bias does not alter the overall conclusion that we can draw from the causal analysis.

**DATA AVAILABILITY**

Components of the PPCG data set can be accessed through different portals in accordance with the required level of data protection for each data type. The main data constituents, and respective modes of access, are listed in detail in the companion manuscript by GM Jakobsdottir[19].

**CODE AVAILABILITY**

Codes are available to reviewers.
Open-source repository will be made available at the time of publication.

**ACKNOWLEDGEMENTS**

We extend our gratitude to all men diagnosed with prostate cancer, as well as the other participants who contributed their time and samples to the consortium. We also acknowledge the dedicated efforts of the research staff across the participating groups, whose careful curation of samples and follow-up data made this work possible. Special thanks go to Alison Thwaites for her tireless coordination and support of the Pan Prostate Cancer Group. A.F.-S. and G.M. are based at the Centro Nacional de Investigaciones Oncológicas (CNIO), which is funded by the Instituto de Salud Carlos III and recognized by the Spanish Ministry of Science and Innovation (MCIN/AEI/10.13039/501100011033) as a 'Severo Ochoa' Centre of Excellence (ref. CEX2019-000891-S). Both A.F.-S. and G.M. also received support from Spanish Ministry of Science and Innovation grants PID2019-111356RA-I00 and PID2023-151298OB-I00 (MCIN/AEI/10.13039/501100011033). Additionally, A.F.-S. was awarded a fellowship from La Caixa Foundation (ID 100010434; LCF/BQ/DR21/11880009). V.J.G. acknowledges infrastructure backing from the NIHR Cambridge Biomedical Research Centre (BRC-1215–20014). V.M.H. received support from the Petre Foundation via the University of Sydney Foundation (Australia). H.H.H. is supported by project grants from the Canadian Institutes of Health Research (CIHR) (142246, 152863, 152864, 159567 and 438793).

This work was also funded by NHMRC project grants 1104010 (C.M.H., N.M.C.) and 1047581 (C.M.H., N.M.C.), as well as through a federal grant from the Australian Department of Health and Ageing awarded to the Epworth Cancer Centre, Epworth Hospital (N.M.C., C.M.H.). We acknowledge further financial support from Australian Prostate Cancer Research and the University of Melbourne, Australia. M.L. received funding from National Cancer Institute grants P50CA211024, P01CA265768, R01 CA259200, from the U.S.A. Department of Defense (DoD) grants PC160357 and PC200390, as well as from the Prostate Cancer Foundation (22CHAL05). Additional support for SAPCS analytical costs came from the U.S. National Institute of Health (NIH) National Cancer Institute (NCI) Award R01CA285772-01 and a U.S. Prostate Cancer Foundation (PCF) Challenge Award (2023CHAL4150). Genomic sequencing and investigation of Southern African Prostate Cancer Study (SAPCS) data received funding from the U.S. Congressionally Directed Medical Research Programs (CDMRP) Prostate Cancer Research Program (PCRP), which included an Idea Development Award (PC200390, TARGET Africa) and HEROIC Consortium Awards (PC210168 and PC230673, HEROIC PCaPH Africa1K). R.M. and A.T.P. are supported by The Lorenzo and Pamela Galli Medical Research Trust, and A.T.P. also holds an Investigator Grant (2026643) from the National Health and Medical Research Council (NHMRC). B.P. is the recipient of a Victorian Health and Medical Research Fellowship awarded by the Victoria State Government, Australia. K.D.S. is funded by The Novo Nordisk Foundation (grant nos. NNF20OC0059410, NNF21OC0071712), The Danish Cancer Society (grant no. R352-A20573), and Independent Research Fund Denmark (grant no. 9039-00084B). J.R. acknowledges support from a CIHR Project Grant (grant no. PJT-162410) and an Investigator Award from the Ontario Institute for Cancer Research (OICR), which is itself funded by the Government of Ontario. Work at the University of Konstanz was supported by the university and an Exploration Grant from the Boehringer Ingelheim Foundation to A.J.G. J.W. received grants from the Danish Cancer Society (#R147-A9843, #R374-A22518), the Danish Council for Independent Research (#8020-00282, #3101-00177A), the Novo Nordisk Foundation

(#NNF200C0060141), and Sygeforsikringen Danmark (#2022-0198). A.L. is supported by Cancer Research UK (C57899/A25812), The John Fell Fund (0012782), the Health Technology Assessment (NIHR 131233), and the John Black Charitable Foundation (TRANSLATE Trial-linked biobank). Y-J.L. receives funding from Orchid, Prostate Cancer Research UK & Movember (MA-CT20-011, RIA22-ST2-006) and Cancer Research UK (C16420/A18066). A full list of funding organizations for the Pan Prostate Cancer Group is provided in a companion manuscript[19].

**COMPETING INTERESTS**

The authors declare no competing interests.

**SUPPLEMENTARY MATERIAL**

**Supplementary Files**
Supplementary File 1. Full list of Pan Prostate Cancer Group members and their affiliations.

AUTHOR CONTRIBUTIONS

E.E. developed the model and carried out the computations. D.W. conceived and supervised the study. A.S. developed the code for AR dysregulation data, V.H., A.Z., C.W., S.B., R.G.B., M.N.B., B.B., A.B., G.C.-T., K.C.L.C., C.S.C., N.M.C., O.C., R.A.E., F.F., C.G., A.G., E.G.G., V.J.G., A.J.G., A.H., V.M.H., H.H.H., C.M.H., G.M.M.J., C.-h.J., F.K., Z.K.-J., P.L., G.L., M.L., P.L., A.P., D.P., B.P., T.R., J.R., B.R., T.S., K.D.S., S.U., J.W., Y.X., P.P.C.G., A.G.L., D.C.W., D.S.B., D.J.W. and all other members of Pan Prostate Cancer Group provided access to data and/or contributed to gathering, processing and curating data. C.S.C, D.C.W, R.A.E., D.B and Z.K.-J. made a substantial contribution to the organisation and conduct of the study and critiqued the output for important intellectual content. A full list of Pan Prostate Cancer Group members can be found in Supplementary File 1. E.E. and D.W. wrote the manuscript. All authors had access to all data in the study. All authors contributed to the review and the editing of the manuscript. All authors approved the manuscript before submission.